\documentclass{article} 
\usepackage{nips14submit_e,times}
\usepackage{hyperref}
\usepackage{url}
\usepackage[pdftex]{graphicx}
\usepackage{float}
\usepackage{subfig,caption}
\usepackage{verbatim,amsmath}
\usepackage{onimage}
\usepackage{chngcntr}
\counterwithin*{section}{part}

\title{Pixels to Voxels: \\Modeling Visual Representation in the Human Brain}

\author{Pulkit Agrawal$^{1}$, Dustin Stansbury$^{2}$, Jitendra Malik$^{1}$, Jack L. Gallant$^{2,3,4}$
\thanks{$^{1}$Department of Electrical Engineering and Computer Science, $^{2}$ Helen Wills Neuroscience Institute, $^{3}$ Department of Psychology, $^{4}$ Program in Bioengineering} \\
University of California Berkeley \\
\texttt{\{pulkitag,stan\_s\_bury,malik,gallant\}@berkeley.edu} \\
}

\nipsfinalcopy 

\begin{document}
\part*{}

\maketitle

\begin{abstract}

The human brain is adept at solving difficult high-level visual processing problems such as image interpretation and object recognition in natural scenes. Over the past few years neuroscientists have made remarkable progress in understanding how the human brain represents categories of objects and actions in natural scenes \cite{Alex,Dustin}. However, all current models of high-level human vision operate on hand annotated images in which the objects and actions have been assigned semantic tags by a human operator. No current models can account for high-level visual function directly in terms of low-level visual input (i.e., pixels). To overcome this fundamental limitation we sought to develop a new class of models that can predict human brain activity directly from low-level visual input (i.e., pixels). We explored two classes of models drawn from computer vision and machine learning.   The first class of models was based on Fisher Vectors (FV) \cite{Fisher} and the second was based on Convolutional Neural Networks (ConvNets) \cite{Lecun89, Kriz}. We find that both classes of models accurately predict brain activity in high-level visual areas, directly from pixels and without the need for any semantic tags or hand annotation of images. This is the first time that such a mapping has been obtained. The fit models provide a new platform for exploring the functional principles of human vision, and they show that modern methods of computer vision and machine learning provide important tools for characterizing brain function.

\end{abstract}

\section{Introduction}

\label{sec:intro}

A great mystery in neuroscience is understanding how the brain effortlessly performs high-level visual tasks such as object recognition and scene understanding. Answering this question is of great scientific importance, not only for neuroscience, but also for developing better computer vision algorithms. One common method for investigating vision is to use functional magnetic resonance imaging (fMRI) to measure brain activity from human subjects passively viewing natural images \cite{fMRI_general}. Many experiments contrast brain activity elicited by specific images categories \cite{localizers}. This functional localizer approach has been used to identify many regions of interest (ROIs) in the visual pathway that appear to represent high-level semantic information \cite{rois_general}. Some ROIs  appear to represent the presence of animate features such as body parts and faces: the extrastriate body area (EBA; \cite{EBA}), occipital face area (OFA; \cite{OFA}), and the fusiform face area (FFA; \cite{FFA}). Others appear to represent information contained in natural scenes: the occipital place area (OPA; \cite{OPA}), the parahippocampal place area (PPA; \cite{PPA}), and the retrosplenial cortex (RSC; \cite{RSC}).

Recently a more powerful approach for investigating visual representation in the human brain has been developed, based on the idea of formulating computational encoding models. Encoding models aim to create a nonlinear mapping between the stimulus and measured brain activity. The encoding model approach is much more sensitive than the conventional functional localizer approach because a single experiment can contain an arbitrary number of categories. Furthermore, the fit encoding models provide quantitative predictions of brain activity for new stimuli that were not used to fit the models. 

One previous study used the encoding model approach to investigate semantic representation in higher visual areas of the human brain. This study used a linearizing feature space to mediate between the visual stimuli and measured brain activity. The features were obtained by annotating images by hand, using binary vectors that indicated the presence or absence of specific semantic categories \cite{Tolga}. These feature vectors were then regressed onto brain activity using regularized linear regression. 

One drawback of the approach used in the earlier study is that each image was annotated by hand by a human. This hand annotation process inevitably introduces subjective selection and interpretation bias into the labels, and this inevitably biases the fit encoding models. For example, an image of a rose can be labeled as a rose, flower, plant, or shrub, depending on the intuitions of the labeler. Another significant drawback to hand annotation is that it is slow. The speed of hand annotation inevitably constrains the number of labeled images that can be used as input to the encoding model. This will inevitably limit the space of encoding models that can be explored, and so will likely produce a suboptimal model of the brain.

Although the previous semantic encoding model provided good predictions of human brain activity in higher visual areas, the requirement of hand annotation is unsatisfying. A fully satisfying model of human vision should predict activity across the entire visual hierarchy directly from pixels, without the need for any human intervention. This requires a means for automatically encoding the semantic content of a scene directly from pixels. Recent breakthroughs in computer vision and machine learning have provided algorithms that can accurately perform many high-level visual tasks, such as object recognition and scene classification, directly from image pixels \cite{Kriz, Rcnn}. Motivated by these results, here we use computer vision and machine learning algorithms to create candidate feature spaces that are used in turn to model brain activity. We focus on two state-of-the-art computer vision algorithms. The first is based on Fisher Vector (FV) encoding of local image descriptors \cite{Fisher, FisherMain}. The second is based on a hierarchical image representation learned by a convolutional neural network (ConvNet) \cite{Lecun89, Kriz}. We use feature spaces discovered by both of these approaches to model brain activity in single voxels distributed across visual cortex. We also compare the performance of these models to previously-published encoding models based on semantic annotations provided by humans \cite{Tolga}.

We show that both the FV and ConvNet models predict human brain activity accurately in high-level visual areas, and that the performance of these models is commensurate with the earlier model based on human annotations. However, the FV and ConvNet models also predict activity accurately in early and intermediate stages of the visual pathway, which the model based on hand annotaction cannot do. The utility of the FV and ConvNet models for predicting responses across the visual hierarchy suggests that they might also be useful for investigating low-and intermediate-level visual processing in the human brain. Finally, we show that the ConvNet model can be used to recover the visual receptive fields from single voxels. This is another unique benefit that cannot be obtained from encoding models based on semantic tags. Taken together, these results clearly demonstrate the power of combining computer vision and machine learning methods with the encoding model approach to human vision.. These results also demonstrate the value of methods that bridge computer vision, machine learning and neuroscience.

\section{Methods}

\label{sec:method}

\begin{figure}

\centering

\subfloat{\begin{tikzonimage}[width=1.00\linewidth]{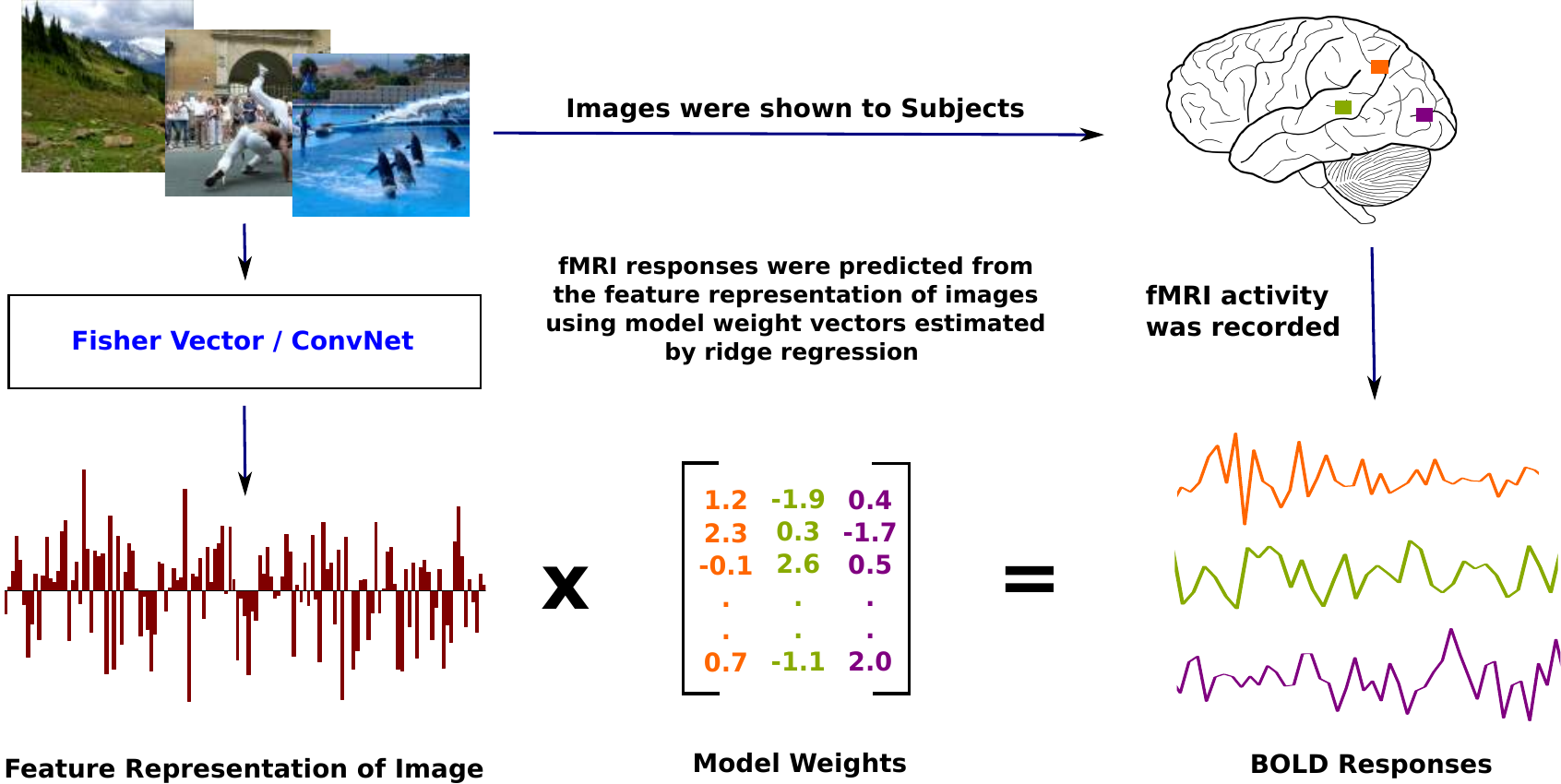} \end{tikzonimage} }  

\caption{Description of the encoding model approach. Regularized linear regression was used to regress the features (bottom left) extracted from the 1260 images in the estimation set (top left) onto brain activity recorded from each voxel (bottom right). The resulting encoding model weights (bottom center) can be interpreted as a tuning curve for the image features. In this work the features were obtained using Fisher vectors (FV), convolutional networks (ConvNet), and a 19 category (19-Cat) model created by hand annotation of the images. Model performance was evaluated by using the fit models to voxel activity elicited by the 126 validation images.}

\label{fig:model}

\end{figure}

The source data for this study were functional magnetic resonance imaging (fMRI) recordings of human brain activity (specifically, the blood-oxygenation-level-dependent (BOLD) signal), recorded continuously while two subjects passively viewed a series of static photos of color natural scenes \cite{Dustin}. The original study measured brain activity elicited by 1260 images shown twice each, and another set of 126 images shown 12 times each. Activity was measured in ~100,000 voxels (i.e., volumetric pixels) located in the cerebral cortex of each subject. We used FV and ConvNet to construct a separate encoding model for each voxel that mapped optimally from the pixels in each estimation image into brain activity evoked by that image (Figure \ref{fig:model}). We then evaluated predictions of the fit models using brain activity evoked by the validation images.

\begin{figure}[t!]
\centering
\subfloat{\begin{tikzonimage}[width=1.00\linewidth]{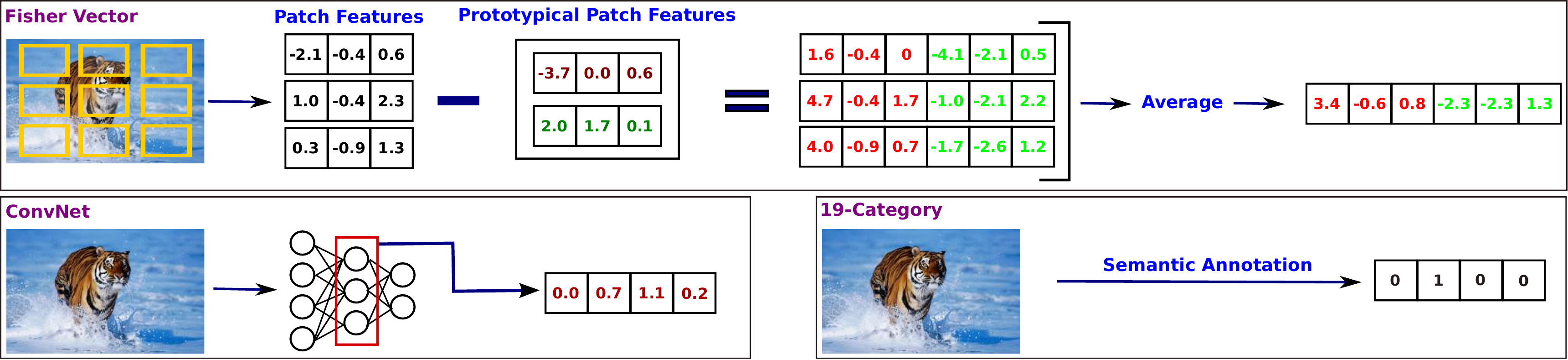}
\end{tikzonimage} }  

\caption{Summary of the three classes of encoding models used in this work. Each model aims to describe the transformation between pixels and measured voxel activity in terms of a projection through an intermediate linearizing feature space. FV (top): Patches were first sampled from each image using a uniform spatial grid (yellow boxes), and SIFT features were calculated for each patch. The vector distance between these SIFT features and each member of a large set of prototypical patch features were then computed. The concatenated mean vector distance of all patches was then used as the FV feature representation for the image. ConvNet (bottom left): ConvNet was fit to a large sample of natural images using standard methods. Then it was used to find features in the 1260 estimation images. Each image was represented by the feed-forward feature activations at one of the 7 layers of the ConvNet. The appropriate layer was selected independently for each voxel. 19-Cat (bottom right): The 19Cat model consisted of a set of high-level semantic image categories annotated by hand \cite{Tolga}. Each image was represented as a binary vector indicating the presence (1) or absence (0) of each category.}

\label{fig:features}

\end{figure}

\subsection{Constructing Encoding Models}

An encoding model consists of a feature space that provides a linearizing transformation between the stimulus images and measured brain activity. Here we constructed three different feature spaces by projecting images in the estimation set \cite{EncfMRI} through FV (Section \ref{subsec:fisher}), ConvNets (sec \ref{subsec:conv}) and the 19-Cat space (Sections \ref{subsec:19-Cat}--\ref{subsec:conv}). Next, for every voxel, we used regularized linear regression to find a set of weights that predicted voxel activity from the feature-space representations of each image. (A single regularization parameter was chosen for all voxels, using five-fold cross-validation.) The accuracy of each encoding model for each voxel was expressed as the correlation ($r$) between predicted and observed voxel activity, using the validation set reserved for this purpose. The explained variance in each voxel's responses was calculated as the square of correlation coefficient $(r^2)$ \cite{David_predicting}. Prediction accuracy was deemed statistically significant if the correlation exceeded $r = 0.33$ ($p < 0.0001$) (for details, see Supplementary Material).

\subsection{Fisher-Vector (FV) Feature Representation}

\label{subsec:fisher}

The FV encoding model used a feature space derived from high-order edge statistics of natural image patches. To learn the feature space, a dictionary of $K$ prototypical image patch features was first learned using a Gaussian Mixture Model (GMM) applied to the SIFT descriptors obtained from thousands of random natural image patches. SIFT features capture the distribution of edge orientation energy in each patch \cite{Sift}. The number of prototypical patches in the dictionary was 64. We chose this value to maximize prediction accuracy of voxel activity, using a portion of the estimation data reserved for this purpose. FV features for each image reflect the vector distance between SIFT features for patches sampled from a multi-scale grid of locations across that image, and the prototypical features learned by the GMM \cite{FisherMain} (see Figure \ref{fig:features} for an illustration).

\subsection{Convolutional Neural Network (ConvNet) Feature Representation}

\label{subsec:conv}

The ConvNet encoding model used a feature space derived from the various layers of ConvNet. The feature space is learned by training ConvNet on the task of image classification \cite{Lecun89},\cite{Kriz}. Here we use the seven-layered ConvNet architecture proposed by \cite{Kriz},\cite{Caffe}. The first five layers are devoted to convolutions (denoted conv-1 through conv-5). The last two layers are fully connected (fc-6, fc-7). We trained the ConvNet on the ImageNet database \cite{Imagenet}, which consists of over 1 million natural images classified into 1000 distinct object categories. The ConvNet features for each image in the estimation set consisted of the feature activations in each layer of the ConvNet. This resulted in 7 possible feature spaces for each image, one corresponding to each layer of the ConvNet. We selected the optimal ConvNet feature space for each voxel by maximizing prediction accuracy of voxel activity, using a portion of the estimation data reserved for this purpose (Figure \ref{fig:features}).

\subsection{19-Category (19-Cat) Feature Representation}

\label{subsec:19-Cat}

The 19-Cat encoding model used a simple feature space that consists of 19-dimensional binary vector indicating the presence (1) or absence (0) of 19 semantic image categories (e.g. "furniture," "vehicle," "water," etc. (see \cite{Tolga}). These categories are likely only a small subset of all those represented in the human brain, but because brain activity measurements are signal limited the 19-Cat model predicts brain activity nearly as well as more complicated semantic models \cite{Tolga}. We used the 19-Cat model as the benchmark for evaluating the FV and ConvNet models developed in this study.

\section{Encoding Model Performance}

\label{sec:predict}

\setlength{\unitlength}{\linewidth}
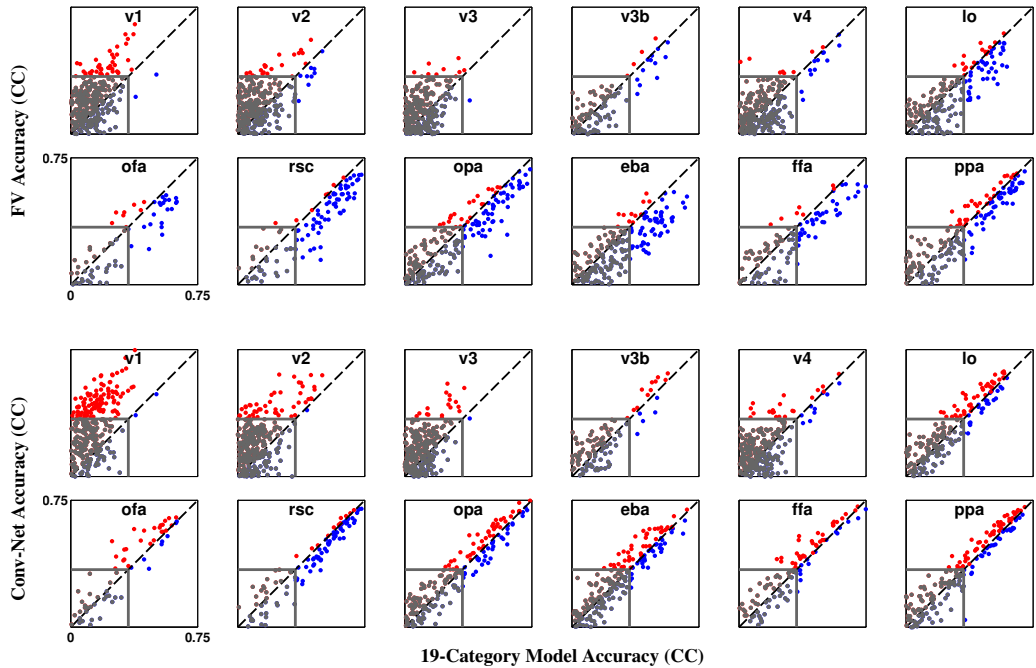
\begin{figure}[t!]
\scalebox{0.80}{
\begin{picture}(0.03,0.3)(0,0)
\put(0,0.1){\rotatebox{90}{\footnotesize{\textbf{FV Accuracy (CC)}}}}
\end{picture}}
\centering
\subfloat{\includegraphics[width=0.95\linewidth]{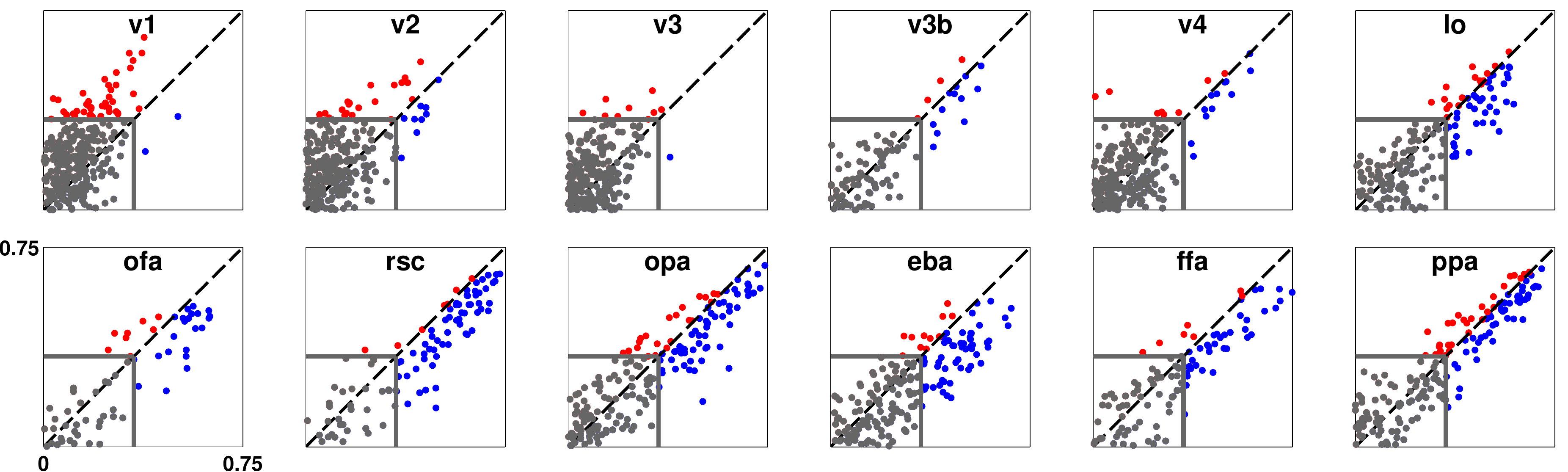} } \vspace{0.01\linewidth}
\vspace{-0.02\linewidth}
\scalebox{0.80}{
\begin{picture}(0.03,0.3)(0,0)
\put(0,0.05){\rotatebox{90}{\footnotesize{\textbf{Conv-Net Accuracy (CC)}}}}
\end{picture}}
\subfloat{\includegraphics[width=0.95\linewidth]{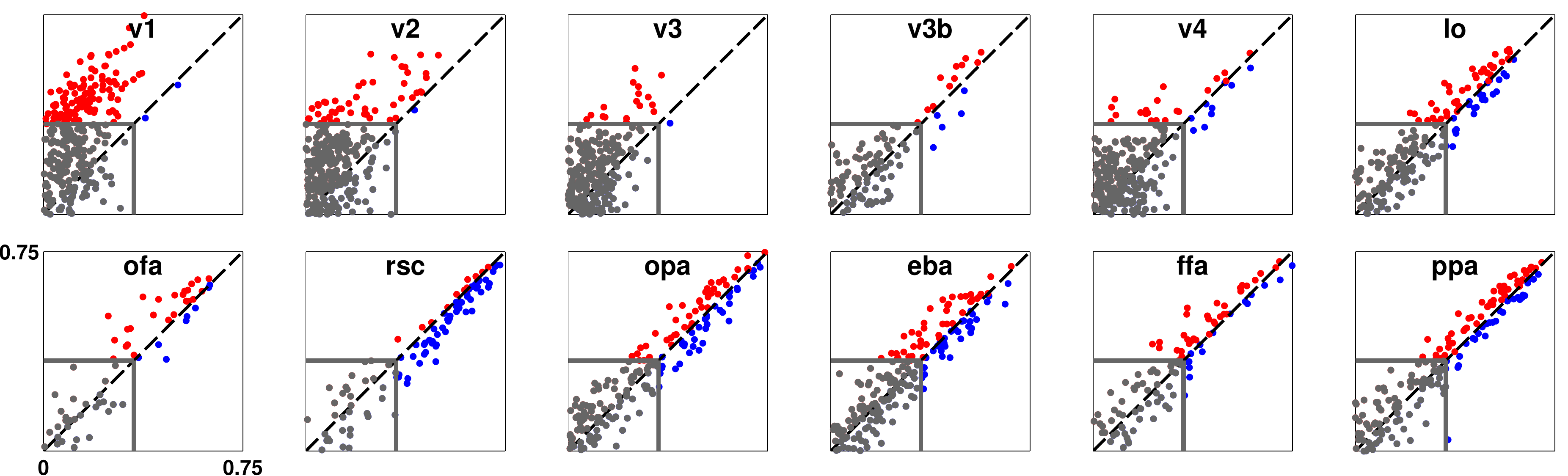} }\vspace{0.01\linewidth}
\scalebox{0.80}{
\begin{picture}(1.0,0.02)(0,0)
\put(0.37,0){\footnotesize{\textbf{19-Category Model Accuracy (CC)}}}
\end{picture}}
\caption{Comparison of predictions of the various models. Each subplot compares predictions of the FV (top) or ConvNet (bottom) encoding model to the 19-Cat model for one visual ROI. Points in each panel represent single voxels. Red dots indicate voxels where predictions of the FV or ConvNet model exceed those of the 19-Cat model, and blue dots indicate voxels best predicted by 19-Cat model. Dots falling along the dashed diagonal line indicate voxels where both models perform equally. Dots in gray fall below the significance threshold ($p <0.0001$) and are indistinguishable from noise. Both the FV and ConvNet models perform better than the 19-Cat model in lower visual areas (V1 and V2) but they are correlated with the 19-Cat model in higher areas. Thus, FV and ConvNet provide feature spaces that can model many visual areas, while the 19-Cat model only model voxels in high-level visual areas.}

\label{fig:scatter_compare}

\end{figure}

\vspace{-4mm}


\begin{figure}[t!]
\subfloat[FV]{\includegraphics[height=0.30\linewidth]{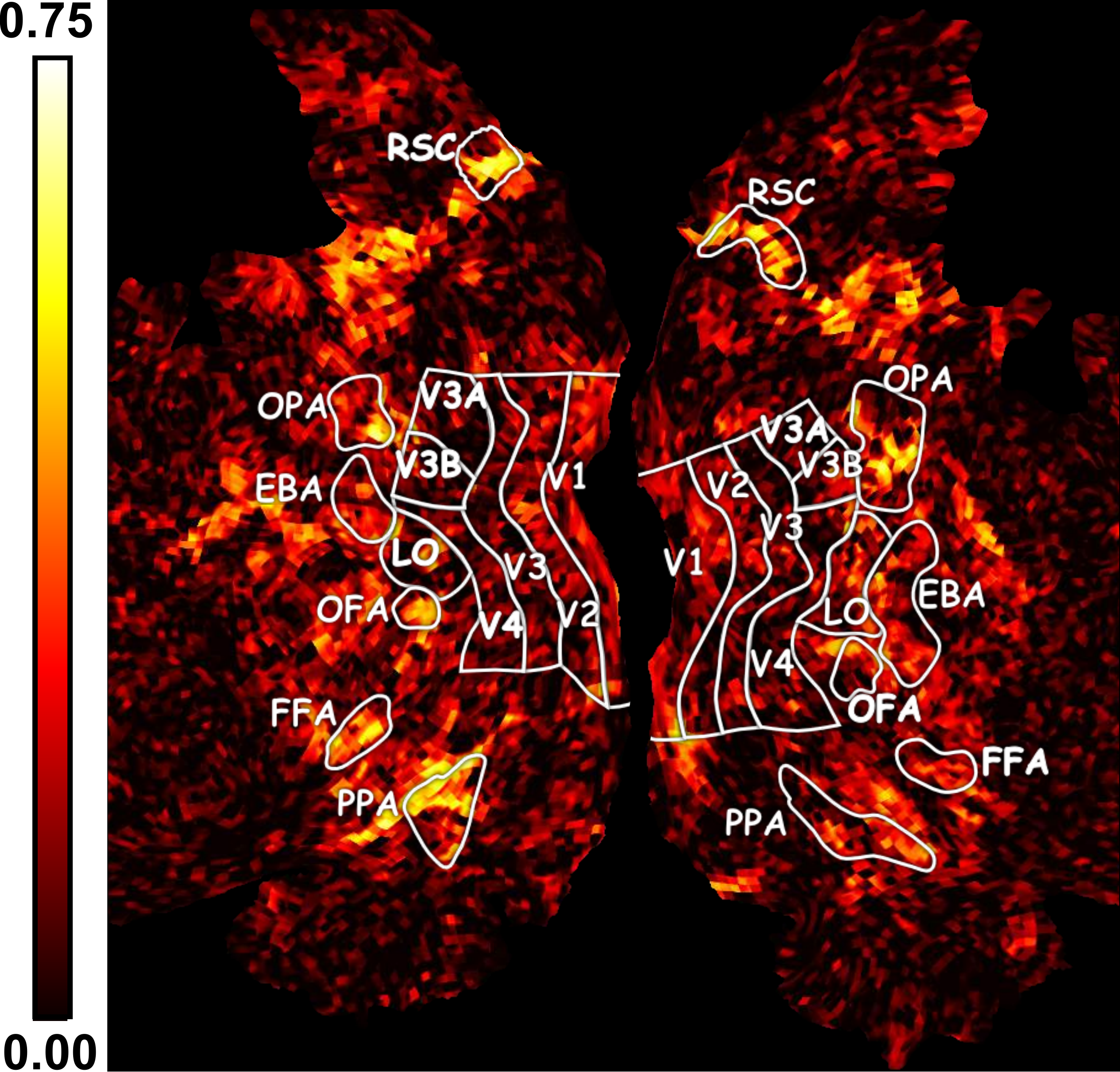} }\hspace{0.01\linewidth}
\subfloat[ConvNet]{\includegraphics[height=0.30\linewidth]{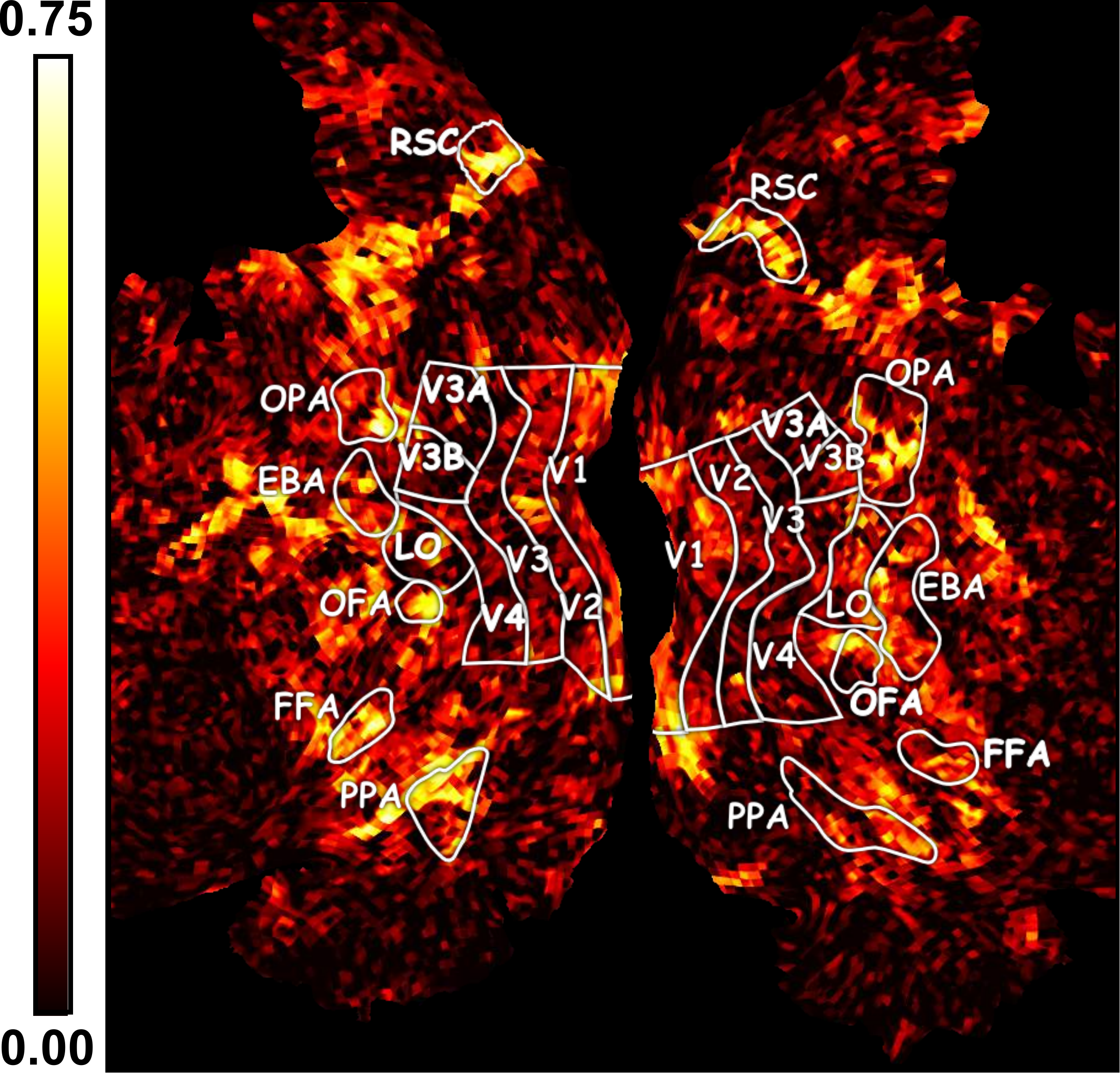}} \hspace{0.01\linewidth}
\subfloat[Flat-Map]{\includegraphics[height=0.30\linewidth]{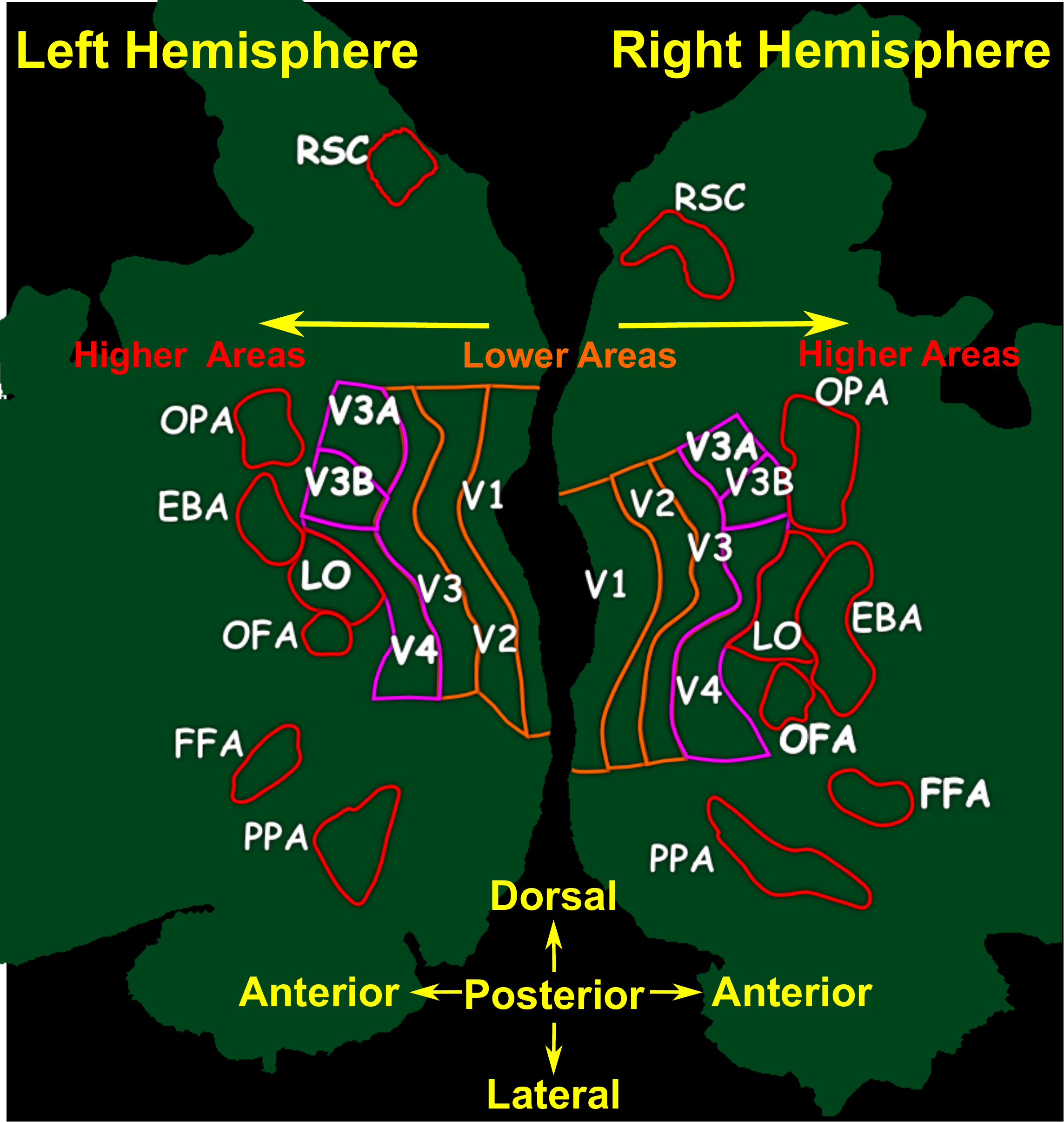}}\\
\subfloat[ConvNet v/s FV]{\includegraphics[height=0.30\linewidth]{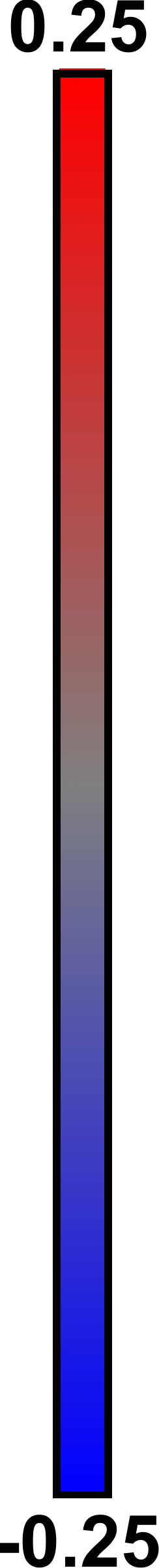}\includegraphics[height=0.30\linewidth]{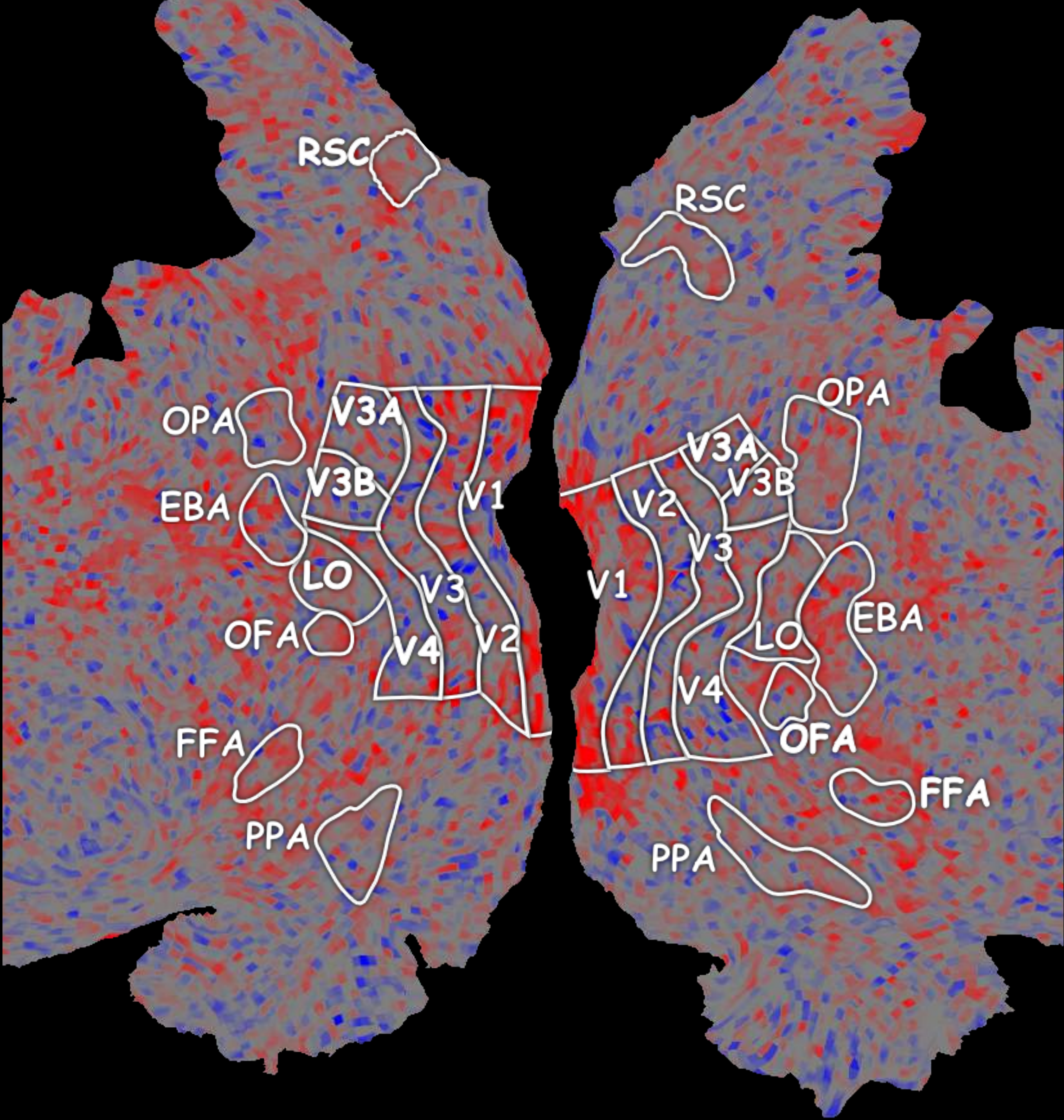}} \hspace{2mm}
\begin{picture}(0.70,0.30)(0,0)
\put(0,0.05){\subfloat[]{\includegraphics[width=0.65\linewidth]{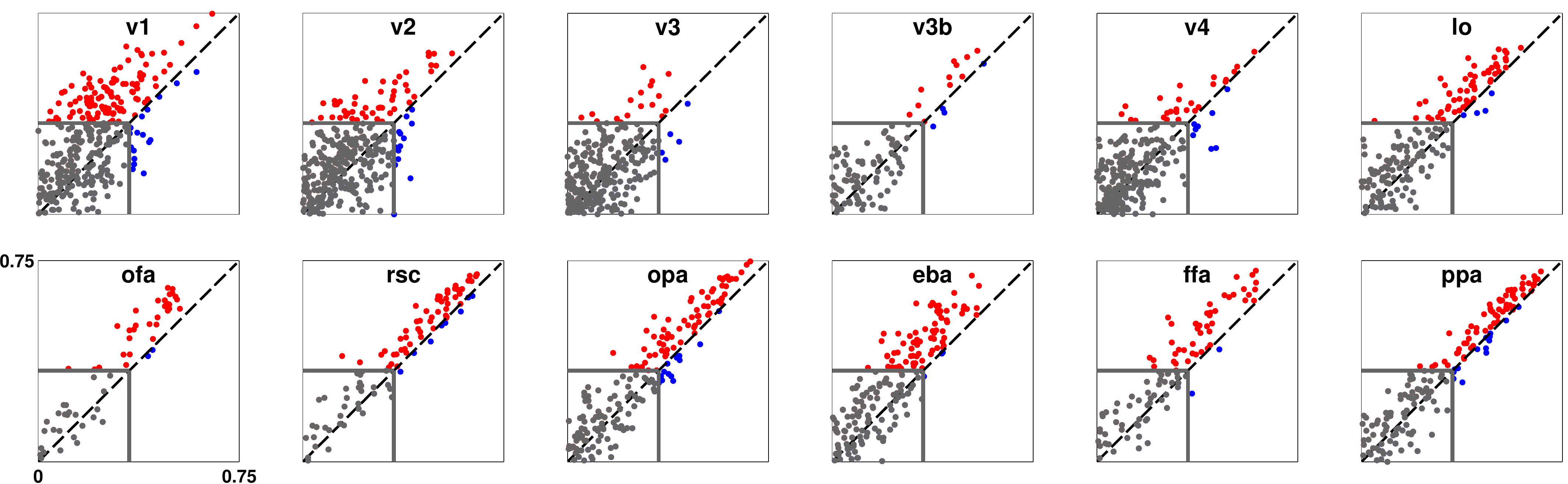}}}
\end{picture}
\caption{Prediction accuracy for the FV and ConvNet models projected onto a flattened map of the visual cortex. Prediction accuracy for the FV model (a) and for the ConvNet model (b). Brighter locations indicate locations where the FV or ConvNet model accurately predicts activity. The Scale bar is shown at left of each panel and the key for identifying brain locations is provided in (c). Panel (d) shows the difference in prediction accuracy between the FV and ConvNet models. Blue locations are best predicted by the FV model. Red locations are best predicted by the ConvNet model. Panel (e) compares predictions of FV and ConvNet models using a format similar to that used in Figure \ref{fig:scatter_compare}. Blue dots indicate voxels best predicted by the FV model and red dots indicate voxels best predicted by the ConvNet model. The results in (d) and (e) show that the ConvNet model outperforms the FV model in early visual areas. However, predictions of the two models are highly correlated in higher visual areas.}
\label{fig:fv_conv}
\end{figure}

In the encoding model framework, the evidence for or against any model is provide by predictions of brain activity in a separate data set reserved for this purpose. Therefore, to determine whether feature spaces derived from computer vision and machine learning are useful for explaining human vision we compared predictions of FV and ConvNet encoding models to the performance of the established 19-Cat model. This comparison is summarized in Figures \ref{fig:scatter_compare} and \ref{fig:fv_conv}. 

For voxels located in lower visual areas, both the FV and ConvNet based models outperform the 19-Cat model. This result makes sense: the feature spaces represented by the FV and ConvNet models incorporate structural information which is absent from the 19-Cat model. Because early visual areas are known to be selective for structural information in natural images, the FV and ConvNet models can predict voxel activity in early areas but the 19-Cat model cannot. For voxels located in higher visual areas,, predictions of the FV and ConvNet models are correlated with those of the 19-Cat model, but the ConvNet model outperforms the 19-Cat model in a few anterior areas (e.g. OFA and EBA). Thus, encoding models based on feature spaces derived using current computer vision or machine learning algorithms predict activity in many visual areas better than models built using hand annotations. 

Figure \ref{fig:fv_conv} shows that the ConvNet model generally provides better predictions than those of the FV model. The first layer of ConvNet learns Gabor-like features whose spatial profiles are similar to V1 receptive fields (\cite{Kriz} \cite{Hubel_v1}), so perhaps this difference is not surprising in area V1. However, the ConvNet model also provides relatively more accurate predictions in intermediate areas like V4 and LO. These areas are believed to be involved in form processing and object segmentation, but the features represented therein are poorly understood. Our results that the intermediate layers of ConvNet may provide a useful set of features for studying visual processing in these intermediate visual areas.

\section{Investigating Voxel Tuning}

\label{sec:selectivity}

\begin{figure}[t!]
\centering
\subfloat{\includegraphics[width=1.00\linewidth]{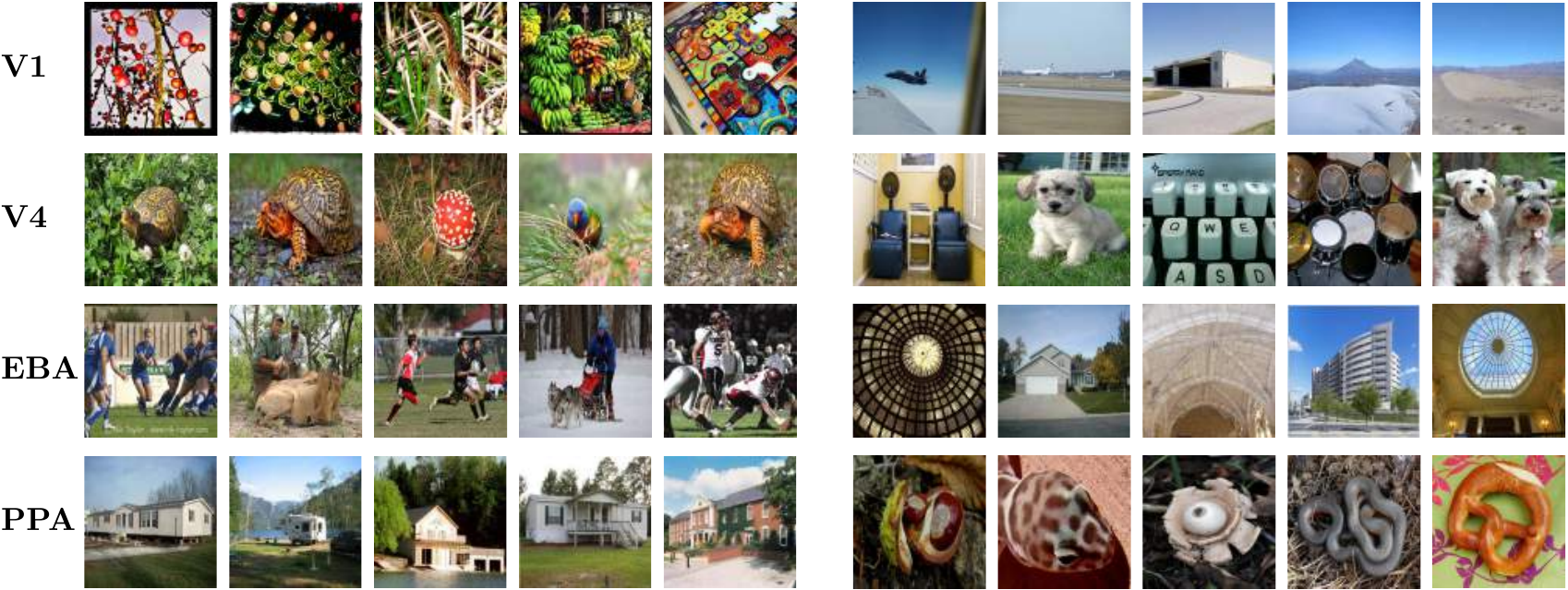}}
\caption{Using the ConvNet model to probe receptive fields of individual voxels. The four rows represent four individual voxels recorded in four ROIs: V1, V4, EBA and PPA. The ConvNet model fit to each voxel was used to filter a set of 170K natural images. The five columns at left show the five images that the ConvNet model for each voxel predicts will most increase activity, and the five columns at right show the images that the model predicts will most decrease activity. The V1 voxel is predicted to increase activity when images consist of high texture and to decrease activity when images contain blue landscapes. The V4 voxel is predicted to increase activity when images contain a orange blob at center and to decrease activity when they contain textures. The EBA voxel is predicted to increase activity when images contain people or animals and to decrease activity when they contain scenes or texture. The PPA voxel is predicted to increase activity when images contain scenes and to decrease activity when they contain texture.}
\label{fig:tuning}
\end{figure}

The previous section demonstrates that the feature spaces provided by FV and ConvNets can be used to predict accurately brain activity in many visual areas. This suggests that we might be able to gain a better understanding of human visual representation by examining the FV and ConvNet models fit to individual voxels. To visualize the features represented in a single voxel we first used the weights of the fit ConvNet model to generate theoretical responses to a large collection of natural images. (We restricted our analysis to ConvNet models here because they generally provide the most accurate predictions.) We then rank-ordered the images according to the responses predicted by the fit model. The top and bottom images within this ranking provide qualitative intuition about the features that are represented by a particular voxel.

Figure \ref{fig:tuning} shows results for several voxels sampled from the V1, V4, EBA, and PPA. Activity in the V1 voxel is predicted to increase when an image consists of high-frequency texture and activity is predicted to decrease it consists of blue low-frequency texture. This pattern of selectivity is reminiscent of contrast-sensitivity \cite{Hubel_v1} reported previously in neurophysiological studies of V1. Activity of the EBA voxel is predicted to increase when an image contains people or animals and to decrease when it contains scenes or texture, while the PPA voxel is predicted to increase activity when images contain scenes and to decrease activity when they contain texture. These patterns of selectivity are consistent with previous reports of these areas \cite{localizers, EBA, PPA}. The V4 voxel is of particular interest because the visual features represented in area V4 are largely unknown. Activity of the V4 voxel is predicted to increase when the center of the image contains an orange blob and to decrease when it contains large-scale texture. This is at least qualitatively consistent with neurophysiological reports that area V4 is selective for curvature and radial patterns \cite{V4_nonpolar}.

\begin{figure}[t!]
\centering
\subfloat{\includegraphics[width=1.00\linewidth]{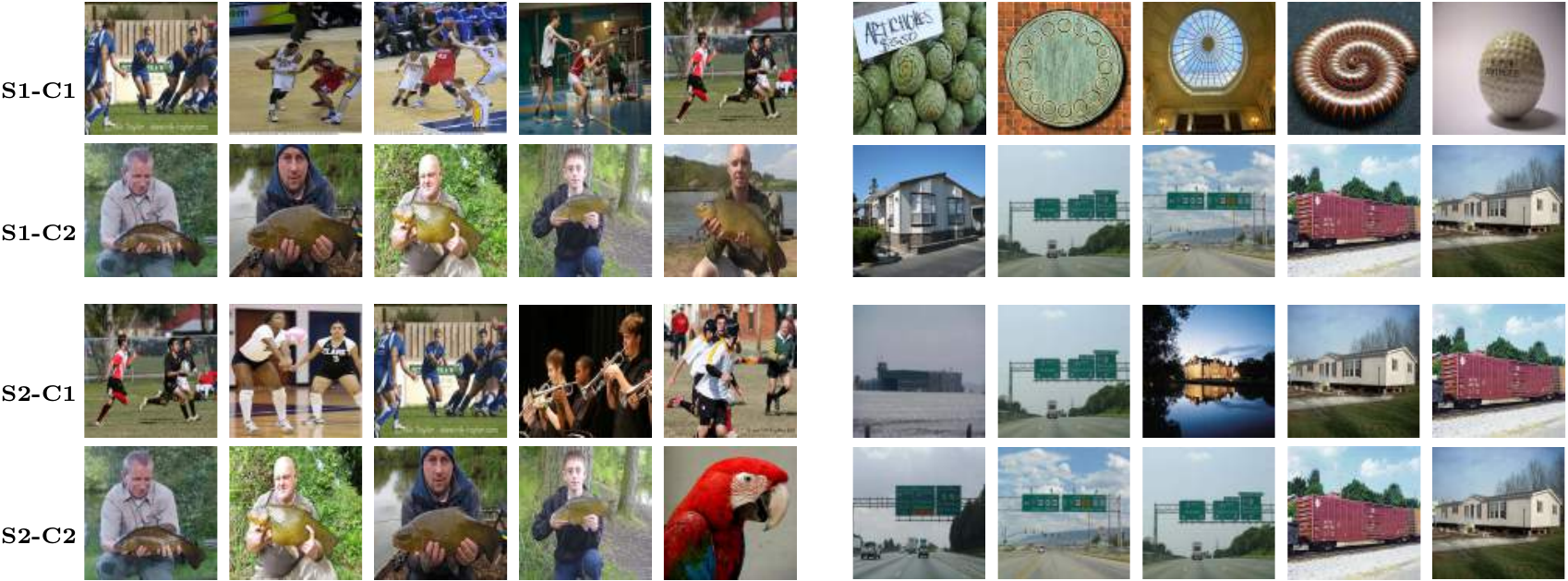}}
\caption{Using the ConvNet model to probe receptive fields of voxel clusters. The four rows represent two EBA clusters in two different human subjects. The average ConvNet model fit to all voxel in each cluster was used to filter a set of 170K natural images. The five columns at left show the five images that the ConvNet model for each cluster predicts will most increase activity in the cluster, and the five columns at right show the images that the model predicts will most decrease activity. Cluster 1 is predicted to increase activity when images contain static pictures of groups of people in action and to decrease activity for rounded textures (S1) or landscape (S2). Cluster 2 is predicted to increase activity when images contain single people and to decrease activtiy when they contain landscapes.}
\label{fig:eba_cluster}
\end{figure}

\begin{figure}[t!]
\centering
{\subfloat{\includegraphics[width=0.20\linewidth]{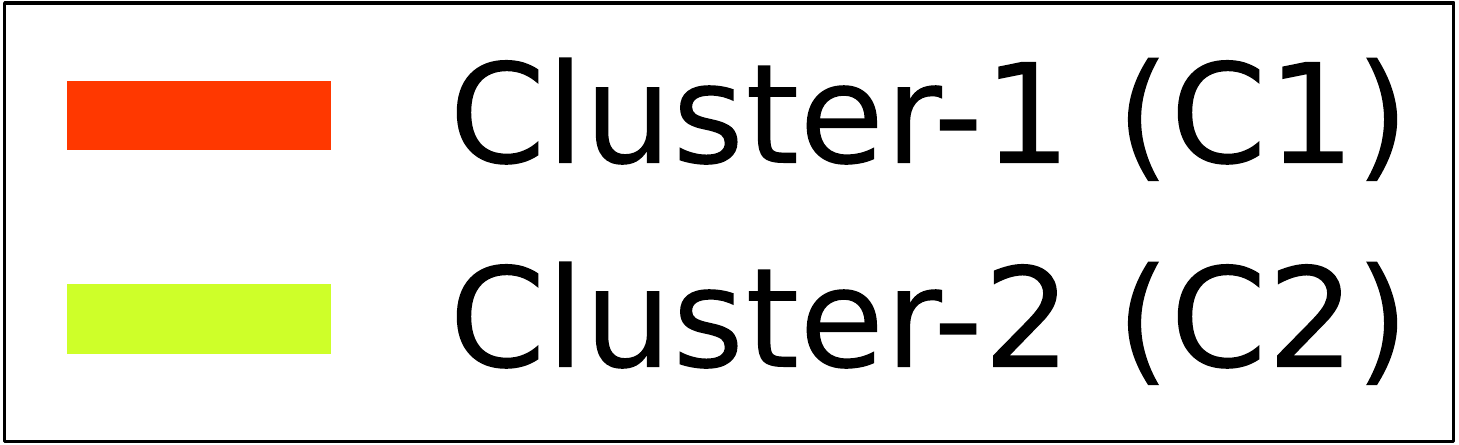}}} \\
\subfloat[S1]{\includegraphics[width=0.39\linewidth]{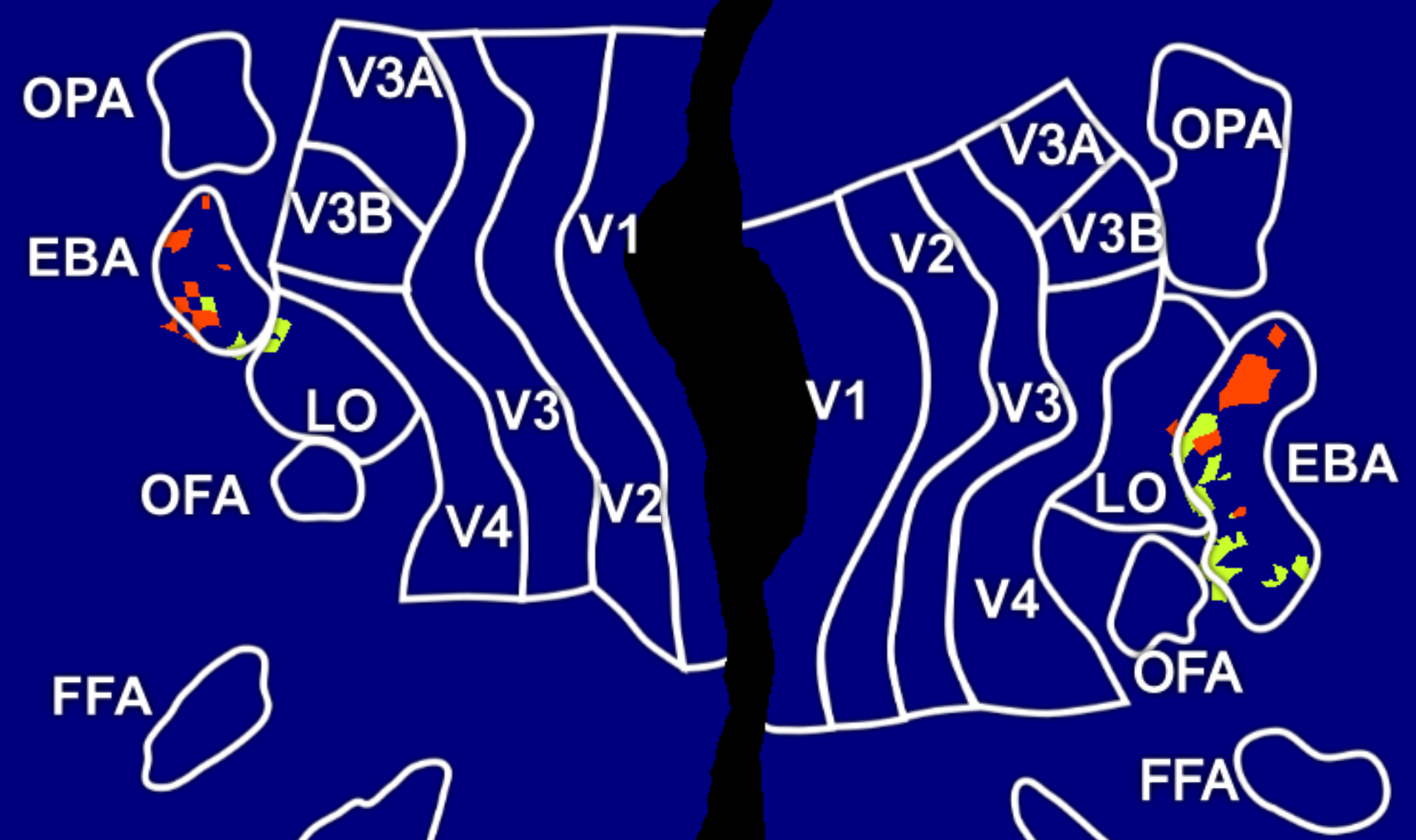}} \hspace{4mm}
\subfloat[S2]{\includegraphics[width=0.43\linewidth]{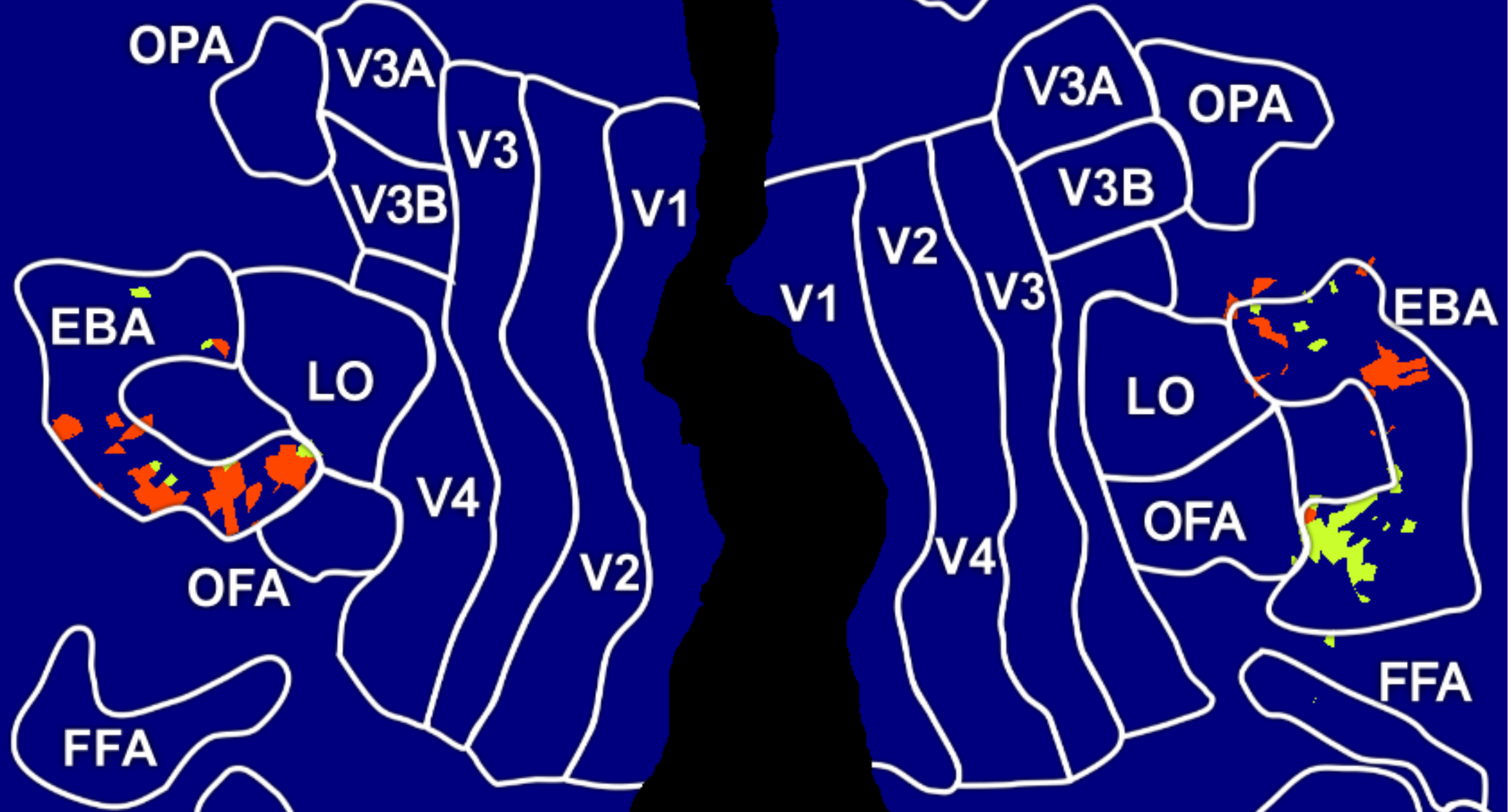}} \hspace{4mm}
\caption{Cortical flat maps showing the location of EBA clusters for the two subjects in this study. The clusters are spatially coherent and they are present at corresponding anatomical locations in both subjects.}
\label{fig:eba-loc}
\end{figure}

The encoding model approach also provides new opportunities for investigating the fine-grained structure of classical ROIs identified in earlier studies \cite{Naselaris_encoding}. As a demonstration of this we performed K-Means clustering of the ConvNet model weights for all of the voxels within area EBA whose activity was predicted significantly. The results revealed two stable clusters of voxels within EBA. One cluster (C1) is predicted to increase activity when images contain full bodies in motion and to decrease when they contain round, up-close objects. The second cluster (C2) is predicted to increase activity when images contain humans and to decrease when they depict outdoor scenes. (Figure \ref{fig:eba_cluster}). Projection of the functional clusters onto cortical flat maps suggests that the clusters are also spatially segregated (Figure \ref{fig:eba-loc}). Thus, this result suggests that EBA contains two distinct functional subdivisions. If this is true, then the average ConvNet model fit to all voxels in C1 should predict the activity of C1 voxels significantly better than the average ConvNet model fit to voxels in C2, and vice-versa. As shown in table \ref{table:eba}, this prediction is confirmed (see Supplemental Material for details). 

\setlength{\tabcolsep}{2pt}
\begin{table}[t!]
\begin{center}
\caption{Test of functional subdivisions within EBA. The test assesses whether the average ConvNet model fit to all voxels in C1 predicts the activity of C1 voxels significantly better than the average ConvNet model fit to voxels in C2, and vice-versa. Data are shown for 2 subjects. A1 refers to the average ConvNet model for cluster C1, and A2 refers to the average ConvNet model for cluster C2. Rows show the variance explained in C1 and C2 by these models. For both subjects, the variance explained by the within-cluster model is significantly higher than variance explained by the model of the other cluster. This confirms that the two clusters have different functional selectivity.}
\label{table:eba}
\scalebox{1}{
\begin{tabular}{|l|cc|cc|}
\hline
 & \multicolumn{2}{c}{Subject-1 (S1)} & \multicolumn{2}{c|}{Subject-2 (S2)} \\
\hline
      & A1 & A2 & A1  & A2\\
\hline
Explained Variance in C1 & $24.9 \pm 1.4\%$ & $19.3 \pm 3.8\%$ &$27.3 \pm 1.0\%$ & $16.2\%\pm 2.0\%$  \\
Explained Variance in C2 & $11.5 \pm 2.3\%$ & $26.2 \pm 1.3\%$ &$14.2 \pm 2.6\%$ & $23.0\% \pm 1.8\%$  \\ 
\hline
\end{tabular}}
\end{center}
\end{table}
\setlength{\tabcolsep}{1.4pt}
\section{Conclusions and Future Directions}
\label{sec:discussion}

In this work we sought to leverage recent advances in computer vision and machine learning to develop encoding models that accurately predict human brain activity evoked by complex natural images. Previous encoding models based on hand annotations of natural images produced good predictions but were unsatisfactory. We investigated two models, one based on FV and one based on ConvNets. We find that these models predict brain activity across many low- and high-level visual areas with an accuracy commensurate with previous models. This is a remarkable result, because these predictions were based entirely on features learned by the FV and ConvNet algorithms and did not require any human annotation. The fact that FV and ConvNet models explain brain activity across visual cortex suggests that human brain is exquisitely tuned to natural scene statistics. 

The ConvNet encoding model provides a powerful new way to investigate visual representation in the human brain. The models fit to individual voxels can be probed in order to visualize the patterns predicted to increase or decrease brain activity. This exercise confirms previous findings and leads to insights about representation in intermediate visual areas. The models can also be used to explore conventional ROIs in more detail. For example, we find that area EBA consists of two functionally and spatially segregated subdivisions. Together, these results demonstrate the power of combining modern methods of computer vision and machine learning with the encoding model approach to fMRI. \\

\noindent\textbf{Acknowledgements} \\ \\
This work was supported by grants to Jack L. Gallant from the National Eye Institute (EY019684), the National Institute of Mental Health (MH66990), and the National Science Foundation Center for the Science of Information (CCF-0939370). Pulkit Agrawal was supported on Fulbright Science and Technology Award. We thank NVIDIA corporation for providing us with GPUs. We thank Alexander G. Huth, Mark Lescroart, Anwar Nunez-Elizalde and Brian Cheung for their helpful discussions and comments.  \\

\noindent\textbf{References}

{\def\section*#1{}

\bibliographystyle{IEEEbib}

{\scriptsize\bibliography{myref}}


}
\clearpage
\part*{}
\setcounter{section}{0}
\setcounter{figure}{0}
\noindent\makebox[\linewidth]{\rule{\linewidth}{3.0pt}}
\begin{center}
\LARGE{\textbf{Supplementary Material}}
\end{center}
\noindent\makebox[\linewidth]{\rule{\linewidth}{0.8pt}}

\section{Testing Overlap in fMRI Stimulus Images and Imagenet Images}
Recall that we trained the ConvNet with a learning database of images and labels taken from the ImageNet ILSVRC12 data set \cite{Imagenet}. Given that we used these features to construct encoding models of fMRI activity, it is thus important to ensure that that the stimuli used in the fMRI experiment do not significantly overlap with the images used to train the ConvNet, as this could result in trivial results.

We determined the amount that the stimulus images overlap with Imagnet by closely following the method proposed by \cite{Rcnn}. The method compares the distance between the GIST\cite{Gist} descriptors of images after resizing them to a size of $32\times32$. We found that there are 5 common images out of a total of 1386 images (1260 training + 126 validation), which is much less than 0.5\% of all the images used for the fMRI experiment. This suggests that there is no substantial overlap between the stimuli and the images used to train the ConvNet.

\section{Computing Significance Values}
In this work we assess encoding model accuracy by calculating the correlation coefficient between the actual responses and responses predicted by the model. When analyzing thousands of models, as is done in this work, it is possible that high correlations can occur by chance. We would thus like to determine a significant level of correlation, when compared to a null distribution obtained by chance. To determine significant correlation, we use a permutation method.

Specifically, for each encoding model, we estimate the prediction accuracy based on 1000 permutations of the validation set responses, giving a distribution of null correlation values. We then take the upper (1-$p$-value)-th percentile of the null distribution as the threshold for significant correlation. The significance value is the probability of observing the correlation between actual and predicted response (without any shuffling) under the null distribution of predictions created by shuffled stimulus-response pairs (p-value).

\section{Encoding Performance for S2}
\begin{figure}[t!]
\centering
\subfloat{\includegraphics[width=1.00\linewidth]{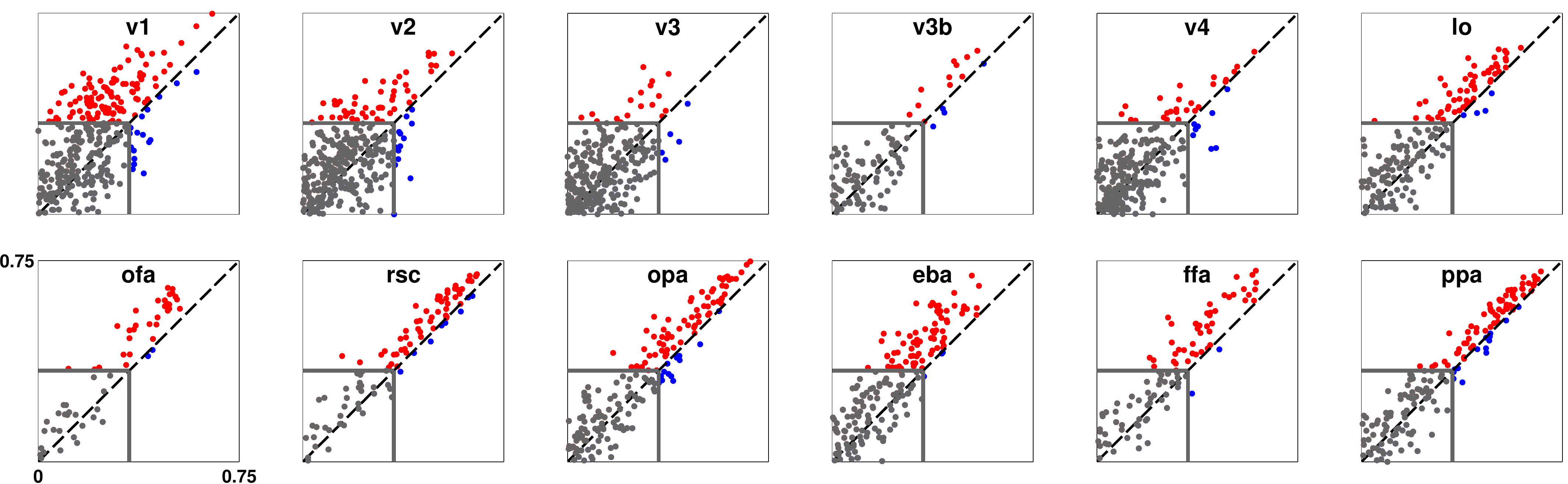}}
\caption{This figure directly compares the predictions of FV and ConvNet using a similar format as Figure \ref{fig:scatter_compare}. However, in this case red dots indicate voxel that are better predicted by the ConvNet models and blue dots indicate voxels that are better predicted by the FV models. In general, the ConvNet models outperform the FV models, as indicated by a majority of red voxels in each plot. However, predictions are very correlated, particularly in late visual areas. This indicates that the two CV-based feature representations encode related information used in high-level visual processing. These results are consistent with the ones found for S1.}
\label{fig:sup_fv_conv}
\end{figure}

Fig\ref{fig:sup_scatter_compare} compares the prediction accuracy of FV and ConvNet models with the 19-Cat model for subject S2. Fig \ref{fig:sup_fv_conv} shows the explicit comparison between the FV and ConvNet models. The results are similar to the ones discussed in the main paper for S1. 
\setlength{\unitlength}{\linewidth}
\begin{figure}[t!]
\scalebox{0.80}{
\begin{picture}(0.03,0.3)(0,0)
\put(0,0.1){\rotatebox{90}{\footnotesize{\textbf{FV Accuracy (CC)}}}}
\end{picture}}
\centering
\subfloat{\includegraphics[width=0.95\linewidth]{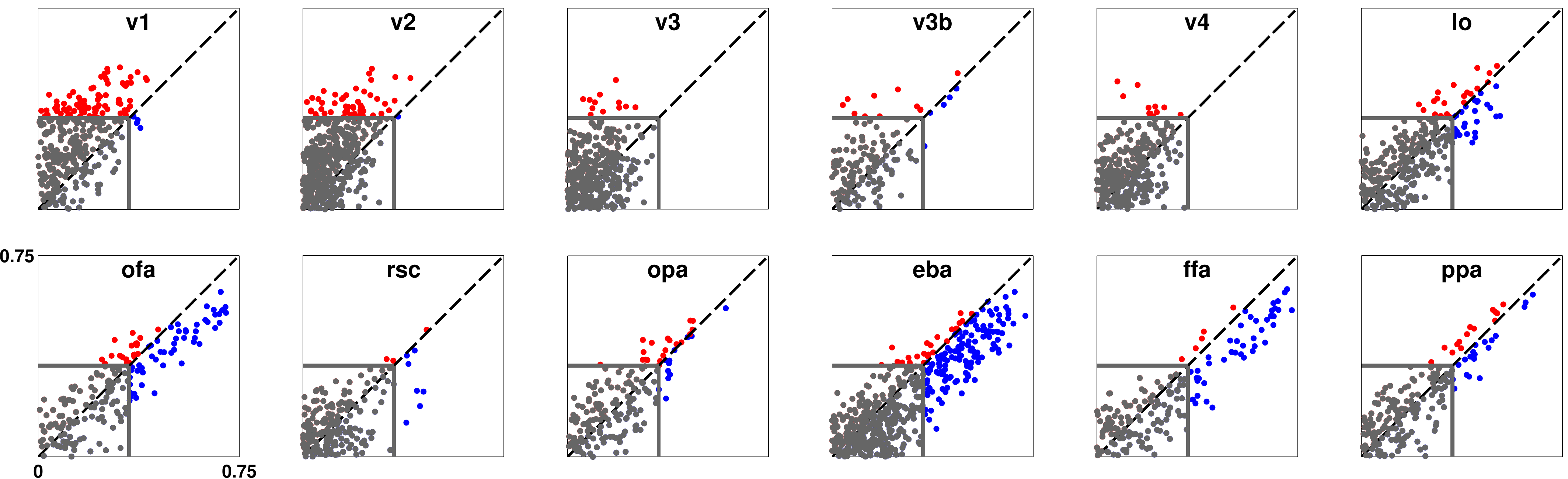} } \vspace{0.01\linewidth}
\vspace{-0.02\linewidth}
\scalebox{0.80}{
\begin{picture}(0.03,0.3)(0,0)
\put(0,0.05){\rotatebox{90}{\footnotesize{\textbf{Conv-Net Accuracy (CC)}}}}
\end{picture}}
\subfloat{\includegraphics[width=0.95\linewidth]{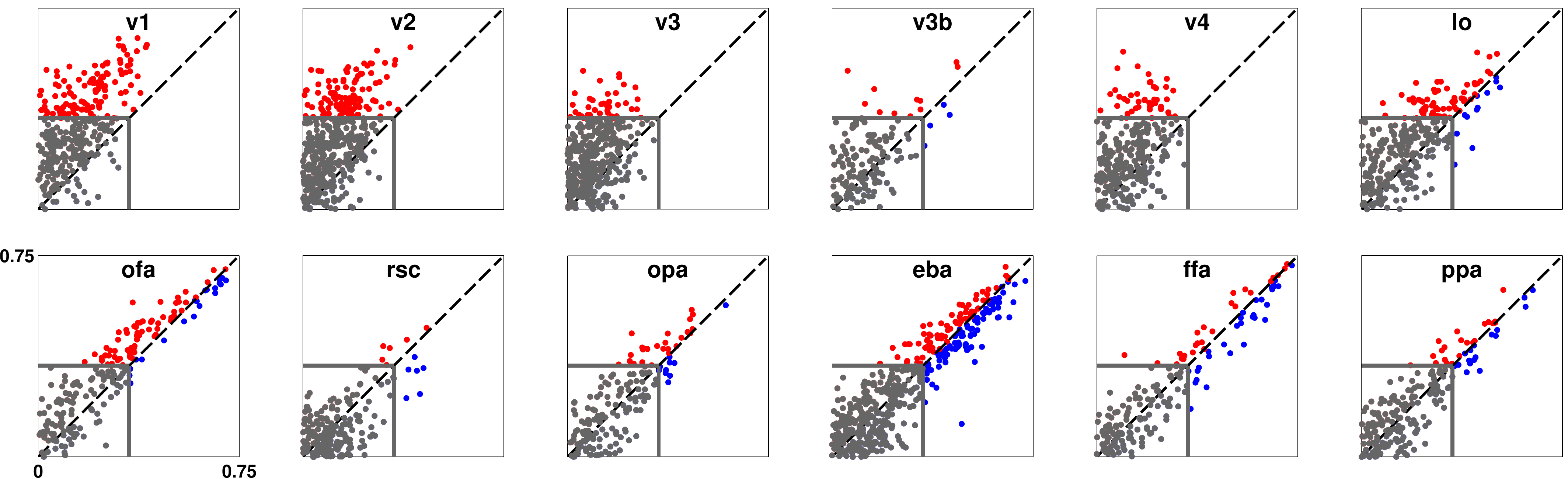} }\vspace{0.01\linewidth}
\scalebox{0.80}{
\begin{picture}(1.0,0.02)(0,0)
\put(0.37,0){\footnotesize{\textbf{19-Category Model Accuracy (CC)}}}
\end{picture}}
\caption{Comparing performance of FV (top) and ConvNet (bottom) models with 19-Cat model in different ROIs of the visual cortex. Voxels shown in red are better predicted by FV/ConvNet model, whereas blue voxels are better predicted by 19-Cat model. The voxels in gray are below the average significance threshold (p-value $ <0.0001$). Note that both FV and ConvNet models perform better in the lower visual areas and are very correlated with the 19-Cat model in the higher visual areas. This indicates that FV/ConvNet models have good prediction in visual areas encoding for semantics.}
\label{fig:sup_scatter_compare}
\end{figure}

\section{ROI Clustering}
\setlength{\unitlength}{\linewidth}
\begin{figure}[t!]
\begin{picture}(0.0,0.3)(0,0)
\put(-0.02,0.06){\rotatebox{90}{\tiny{\textbf{K-Means Energy}}}}
\end{picture}
\centering
\subfloat[K-Means Energy]{\includegraphics[width=0.45\linewidth]{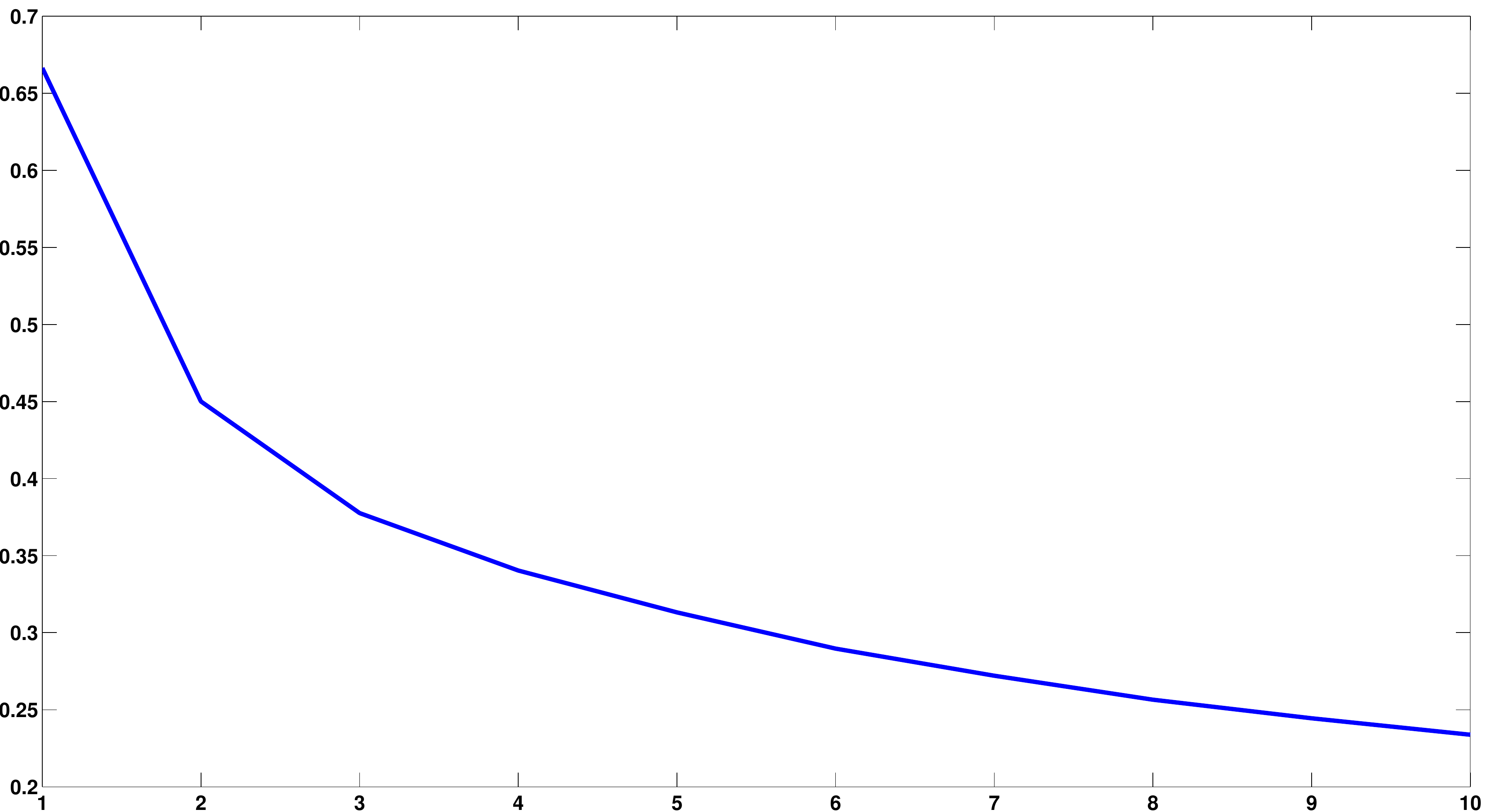}} \hspace{0.02\linewidth}
\begin{picture}(0.0,0.3)(0,0)
\put(0,0.06){\rotatebox{90}{\tiny{\textbf{Relative Entropy}}}}
\end{picture}
\hspace{0.02\linewidth}
\subfloat[Relative Entropy]{\includegraphics[width=0.45\linewidth]{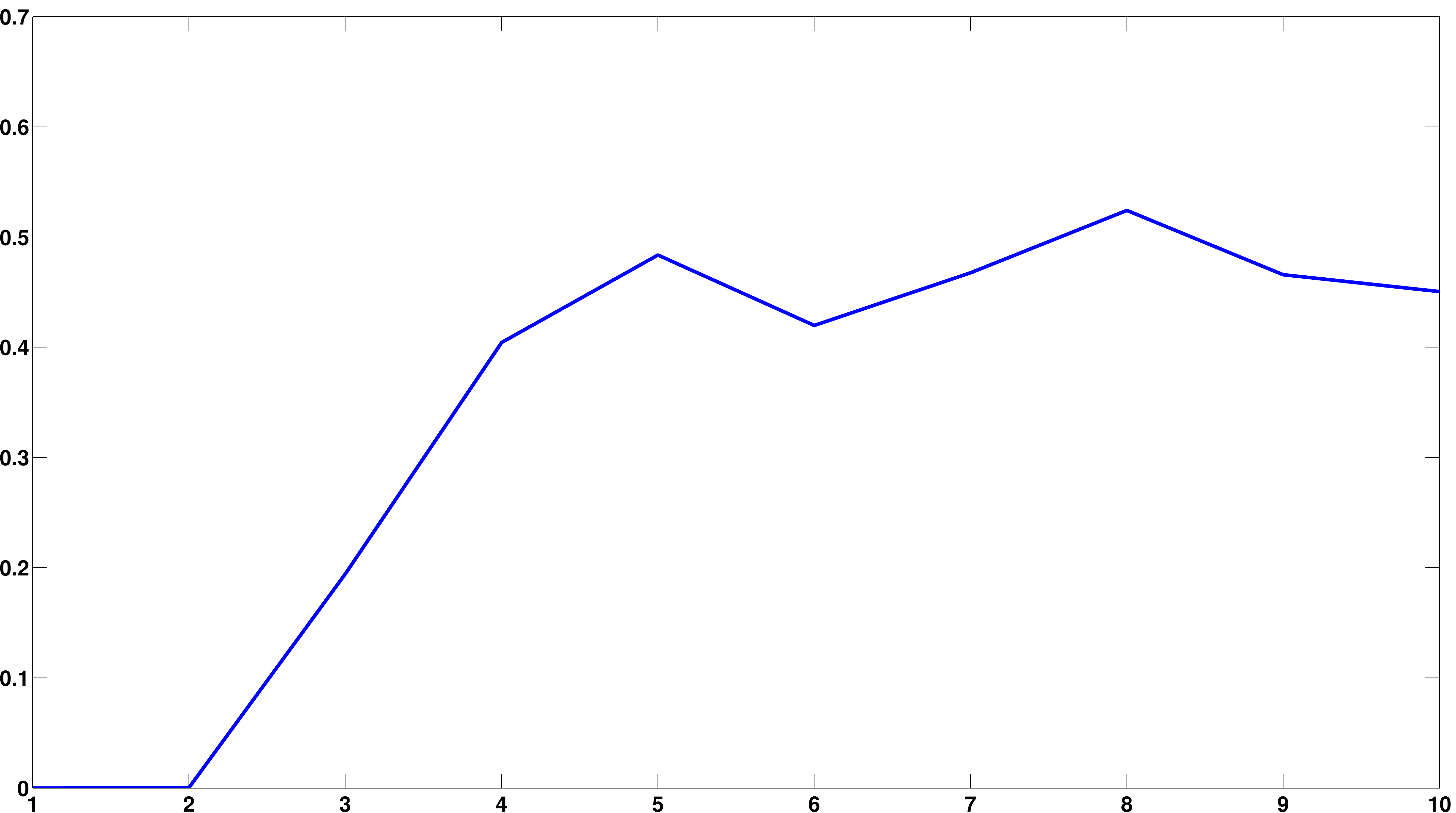} }
\caption{(a) shows residual energy of K-means clustering of model weights of EBA voxels as function of number of clusters (X-axis) for S1. Note, that there is kink at $K=2$. (b) plots the relative entropy of clustering as a function of number of clusters. Low entropy means that the clustering is stable, whereas high entropy indicates unstable clustering. It can be seen that entropy increases substantially for K=3. This indicates that there are atmost two clusters in out data. The plots for S2 follow the same trend.}
\label{fig:sup_eba-cluster}
\end{figure}

In the main text we present results that show the existence of two functional sub-regions in EBA. Our method for identifying these sub-regions using clustering is as follows: First, we trained encoding models for each voxel in EBA based on all layers of the ConvNet and choose the model for each voxel that provides the most accurate predictions on the validation set. For EBA, voxel activity is generally best predicted by models based on layer fc-7 of the ConvNet. This layer defines a 4096-dimensional feature representation of the stimulus. Consequently, the encoding model weights are also 4096-dimensional vectors. We then concatenate the model weights for all EBA voxels and reduced the dimensionality of model weights from 4096 to $\approx$ 100 dimensions using PCA. We then run K-means on PC-reduce model weights.

\subsection{Determining the number of clusters}
The number of clusters ($K$) in the EBA analysis was determined using using two separate methods, each giving the same result. The two methods are described here:
\begin{itemize}
\item{Elbow Method: The intuition behind this method is that one should choose a number of clusters so that adding another cluster does not result in a significant decrease of the objective function optimized by the K-Means algorithm. We refer to the value of this objective function as the energy of the clustering. We calculate the clustering energy for values of $K$ ranging to 1 to 10. Figure \ref{fig:sup_eba-cluster} plots this energy for one subject. It can be seen that the clustering energy does not reduce significantly for values of $K$ greater than 2. This indicates that there are two distinct clusters in the data.}

\item{Entropy based method: The number of clusters is chosen based on the stability of results obtained across 100 bootstrapped repetitions of the clustering method. The intuition is that if the clustering is stable, then across different bootstrap runs, a particular voxel should consistently be assigned to the same cluster. Thus the stability of clustering can is characterized by calculating the entropy across the repeated clustering runs. Low average entropy across all voxels indicates stable clustering across the population. We thus calculate the average entropy across the population of voxels for $K$ = 2 to 10. The results are shown in Figure \ref{fig:sup_eba-cluster} (b). It can be clearly seen that values of $K$ greater than 2 result in high entropy and consequently cluster that are not stable. This suggests that there are two distinct and stable clusters in the data.}
\end{itemize}

\subsection{Average Cluster Model}
The clustering in the PC-space results in an assignment of each voxel to a particular cluster. The average model for each cluster is the mean of model weights of all voxels assigned to that cluster.
\section{Investigating Voxel Tuning}
\subsection{Data-Set Construction}
\label{subsec:data}
The dataset of 170 thousand images used to investigate visual tuning of voxels was constructed by combining images from the Imagenet\cite{Imagenet} validation set and SUN\cite{Sun}, PASCAL \cite{Pascal} image databases. The images span a large variety of natural scenes and objects.

\subsection{Estimating Visual Receptive Fields}
\begin{figure}[t!]
\centering
\subfloat{\includegraphics[width=1.00\linewidth]{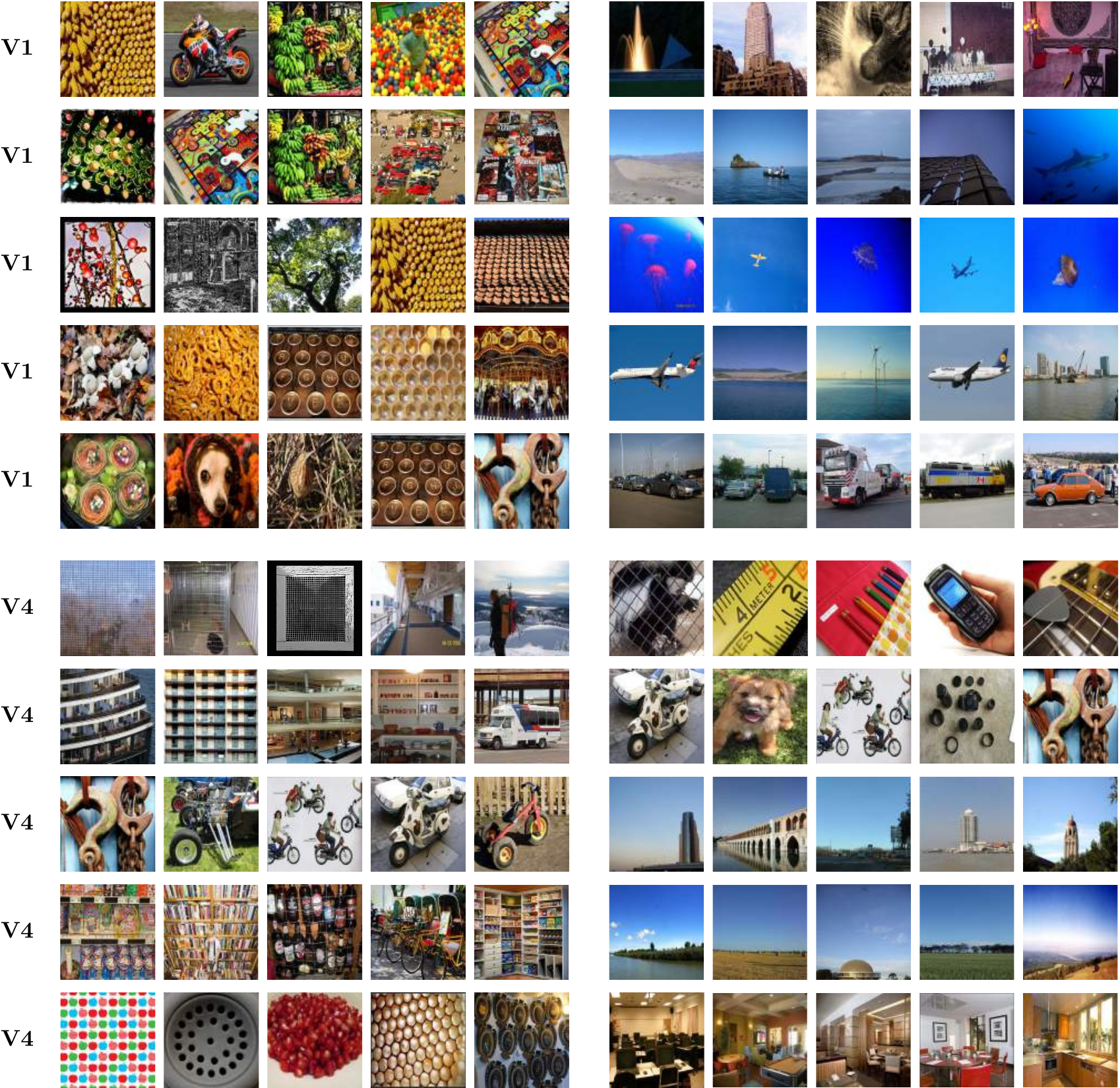}}
\caption{Depiction of visual selectivity of four voxels taken from different ROIs of the visual cortex. The figure shows,  5 voxels each for V1 and V4. The top five and bottom five images (left to right is top to bottom) taken from a ranked list of predicted brain activity by the ConvNet model on a large external dataset of 170K natural images. A voxel is tuned to images which have high predicted response, while it is suppressed by images with low predicted response.  The voxel in V1 seems tuned for images with high texture, whereas V4 voxel seems tuned to horizontal/vertical contours and circular shapes. These results are consistent with previous findings that V4 voxels show tuning for circular objects and concave contours.}
\label{fig:tuning_v1_v4}
\end{figure}

\begin{figure}[t!]
\centering
\subfloat{\includegraphics[width=1.00\linewidth]{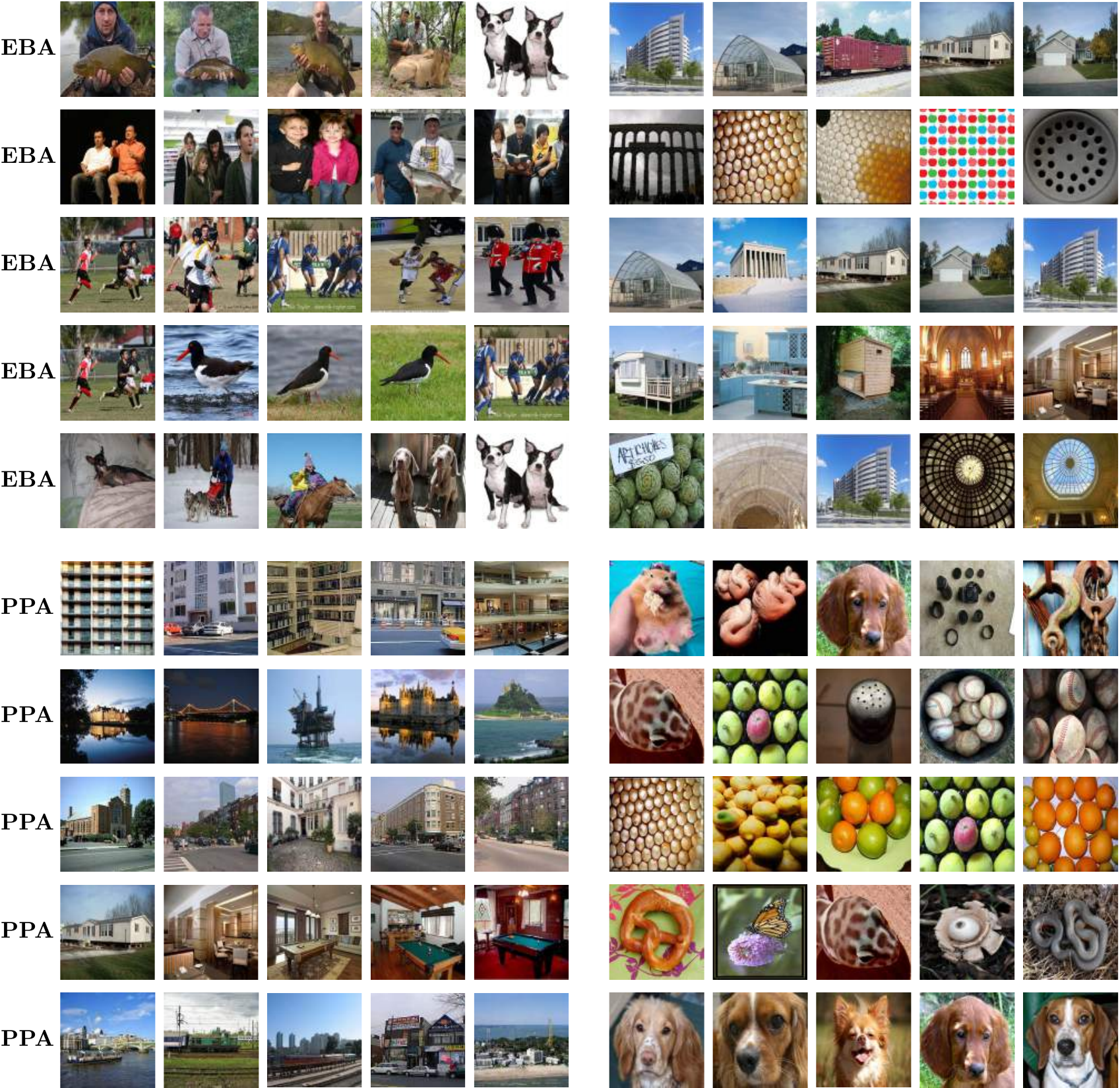}}
\caption{Depiction of visual selectivity of four voxels taken from different ROIs of the visual cortex. The figure shows,  5 voxels each for EBA and PPA. The top five and bottom five images (left to right is top to bottom) taken from a ranked list of predicted brain activity by the ConvNet model on a large external dataset of 170K natural images.A voxel is tuned to images which have high predicted response, while it is suppressed by images with low predicted response.  The EBA voxels seem tuned for static bodies of people, animals and moving bodies. The PPA voxels show a preference for scenes.}
\label{fig:tuning_eba_ppa}
\end{figure}

Recall that for the purpose of comparing the ConvNet model with FV/19-Cat model we estimated the best layer based on the performance of held-out part of the training set. Since, we are no longer interested in comparing performance - but in actually estimating functional/visual receptive fields, we use the model from the layer of the ConvNet which results into best predictions on the validation set of  fMRI dataset to compute   predicted brain activity for the data-set of images described in \ref{subsec:data}. Since, the validation set has 12 repeats for each image instead of 2 in the training set, it results into a higher SNR and consequently a more accurate representation of the brain activity. This choice for visualization simply reduces noise and does not affect any of our conclusion. 

We provide more examples for estimated voxel responses for four ROIs mentioned in the main paper. Fig \ref{fig:tuning_v1_v4} shows tuning for 5 voxels in V1, V4 each whereas Fig \ref{fig:tuning_eba_ppa} shows tuning for 5 voxels in EBA, PPA each. The results are consistent with the ones described in the main paper.

\end{document}